\documentstyle[11pt]{article}
\begin{document}
\title{Duality Symmetry in Momentum Frame}
\author{{Yan-Gang Miao${}^{{\rm a,b},1}$, Harald J. W.
M$\ddot{\rm u}$ller-Kirsten${}^{{\rm a},2}$
and Dae Kil Park${}^{{\rm c,d},3}$}\\
{\small ${}^{\rm a}$ Department of Physics, University of Kaiserslautern,
P.O. Box 3049,}\\
{\small D-67653 Kaiserslautern, Germany}\\
{\small ${}^{\rm b}$ Department of Physics, Xiamen University, Xiamen 361005,}\\
{\small People's Republic of China}\\
{\small ${}^{\rm c}$ Department of Physics, Kyungnam University, Masan 631-701, Korea}\\
{\small ${}^{\rm d}$ Michigan Center for Theoretical Physics, Randall
Laboratory,}\\
{\small Department of Physics, University of Michigan,}\\
{\small Ann Arbor, MI 48109-1120, USA}}
\date{}
\maketitle
\footnotetext[1]{E-mail address: miao@physik.uni-kl.de}
\footnotetext[2]{E-mail address: mueller1@physik.uni-kl.de}
\footnotetext[3]{E-mail address: dkpark@hep.kyungnam.ac.kr}
\vskip 48pt
\begin{center}{\bf Abstract}\end{center}
\baselineskip 22pt
Siegel's action is generalized to the $D=2(p+1)$ ({\em p} even)
dimensional space-time.
The investigation of self-duality of chiral {\em p}-forms is extended to the
momentum frame,
using Siegel's action of chiral bosons in two space-time dimensions and its
generalization in higher dimensions as examples. The whole procedure of the
investigation is realized in the momentum space which relates to the
configuration space through the Fourier transformation of fields.
These actions correspond to non-local Lagrangians in the
momentum frame. The self-duality of them
with
respect to dualization of chiral fields is
uncovered. The relationship between two self-dual
tensors in momentum space, whose similar form appears
in
configuration space,
plays an important role in the calculation, that is, its application realizes
solving
algebraically
an integral equation.
\vskip 24pt
\noindent
PACS number(s): 11.10.Kk

\newpage
\section{Introduction}
Many models of chiral bosons and/or their generalizations to higher (than two)
space-time
dimensions, {\em i.e.}, chiral {\em p}-forms have been proposed [1-12]. Among
them, some [1-7] are non-manifestly space-time covariant, while the others
[8-12] manifestly space-time covariant. Moreover, these chiral {\em p}-form
models have close relationship among one another, especially various dualities
that have been demonstrated [12-14] in detail. Incidentally, the
self-duality that exists beyond chiral {\em p}-form actions has also been
uncovered [15].

As the
investigations of duality symmetries mentioned above are limited only in
configuration space, it may be interesting to extend these
investigations to the momentum space that relates to the configuration space
through the Fourier transformation of fields. It is quite natural to have this
idea because the Fourier transformation plays an important role in
field theory. Perhaps motivated similarly, the duality in
harmonic oscillators obtained by Fourier decomposition was discussed [16] by
considering several simple models, such as the free scalar, Maxwell and
Kalb-Ramond theories, as examples. However, it is quite unsatisfactory to
leave a variety of attractive chiral {\em p}-form models unnoticed. In this
note we re-investigate in momentum space the duality symmetries of various
chiral {\em p}-forms that exist in the configuration space, and we do
find something non-trivial. The non-triviality we mention here means
duality investigation of non-local Lagrangians and algebraic solution of
integral equations.

We choose
Siegel's action as our example. To this end, we have to generalize the action
to the $D=2(p+1)$ dimensional space-time\footnote[1]{Note that the space-time
is twice odd dimensional, {\em i.e.}, {\em p} is
even throughout this paper.}.
The main reason to make this choice
is that after one makes the Fourier transformation to
fields its formulation is non-trivial because of its cubic Lagrange-multiplier
term as shown below. Starting from this formulation, the whole procedure of
investigation is realized in the momentum space.
As a result, the self-duality of Siegel's action with respect to dualization
of chiral fields is uncovered in the momentum frame. In the next section we
deal with the special case in $D=2$ space-time dimensions, and in section 3 we
turn to the general case in $D=2(p+1)$ dimensions. Finally, section 4 is
devoted to a conclusion.

The metric notation we use throughout this note is
\begin{eqnarray}
& & g_{00}=-g_{11}=\cdots=-g_{D-1,D-1}=1,\nonumber\\
& & {\epsilon}^{012{\cdots}D-1}=1.
\end{eqnarray}
Greek letters stand for indices (${\mu},{\nu},{\sigma},{\cdots}=0,1,2,\cdots,
D-1$) in both the configuration and momentum spaces.
\section{Self-duality of the chiral 0-form action in D=2 momentum frame}
We begin with Siegel's action [8] in $D=2$ space-time dimensions,
\begin{equation}
S_c=\int d^2x \left\{-\frac{1}{2}
{\partial}_{\mu}{\phi}(x)
{\partial}^{\mu}{\phi}(x)
+\frac{1}{2}{\lambda}_{{\mu}{\nu}}(x)
[{\partial}^{\mu}{\phi}(x)-{\epsilon}^{{\mu}{\sigma}}
{\partial}_{\sigma}{\phi}(x)]
[{\partial}^{\nu}{\phi}(x)-{\epsilon}^{{\nu}{\rho}}
{\partial}_{\rho}{\phi}(x)]\right\},
\end{equation}
where ${\phi}(x)$ is a scalar field, and ${\lambda}_{{\mu}{\nu}}(x)$ a
symmetric auxiliary second-rank tensor field. Substituting the Fourier
transformations of ${\phi}(x)$ and ${\lambda}_{{\mu}{\nu}}(x)$,
\begin{eqnarray}
{\phi}(x)&=&\frac{1}{2\pi}\int d^2k{e}^{-ik\cdot x}{\phi}(k),\nonumber\\
{\lambda}_{{\mu}{\nu}}(x)&=&
\frac{1}{2\pi}\int d^2k{e}^{-ik\cdot x}{\lambda}_{{\mu}{\nu}}(k),
\end{eqnarray}
into eq.(2), one arrives at the Siegel action in the momentum frame spanned
by $k_{\mu}$,
\begin{eqnarray}
S_m=&-&\frac{1}{2}\int d^2k (-ik_{\mu})(ik^{\mu}){\phi}(k){\phi}(-k)
\nonumber\\
&+&\frac{1}{4\pi}\int d^2k d^2k^{\prime}
{\lambda}_{{\mu}{\nu}}(-k-k^{\prime})
(ik^{\mu}-{\epsilon}
^{{\mu}{\sigma}}ik_{\sigma})
({ik^{\prime}}^{\nu}-{\epsilon}
^{{\nu}{\rho}}ik^{\prime}_{\rho}){\phi}(k){\phi}(k^{\prime}).
\end{eqnarray}
Note that $S_m$ contains quartic integrals over the momentum space because of
the cubic Lagrange-multiplier term in $S_c$, which shows that $S_m$ is
non-trivial. This non-triviality can also be understood as the non-locality
of the corresponding Lagrangian. Therefore, we have to envisage the duality
investigation of non-local Lagrangians.

We investigate the duality property of $S_m$ with respect to dualization of the
field ${\phi}(k)$. By introducing two
independent vector fields in the momentum space, $F_{\mu}(k)$ and $G_{\mu}(k)$, we construct a new
action to replace $S_m$,
\begin{eqnarray}
S_m^{\prime}=&-& \frac{1}{2}\int d^2k F_{\mu}(k)F^{\mu}(-k)\nonumber\\
&+&\frac{1}{4\pi}\int d^2k d^2k^{\prime}
{\lambda}_{{\mu}{\nu}}(-k-k^{\prime})
\left[F^{\mu}(k)-{\epsilon}
^{{\mu}{\sigma}}F_{\sigma}(k)\right]
\left[F^{\nu}(k^{\prime})-{\epsilon}
^{{\nu}{\rho}}F_{\rho}(k^{\prime})\right]\nonumber\\
&+&\int d^2k G^{\mu}(-k)\left[F_{\mu}(k)+ik_{\mu}{\phi}(k)\right],
\end{eqnarray}
where the third term is nothing but the {\em k}-space formulation of
$\int d^2x G^{\mu}(x)[F_{\mu}(x)-{\partial}_{\mu}{\phi}(x)]$. Variation of
$S_m^{\prime}$ with respect to $G^{\mu}(-k)$ gives
\begin{equation}
F_{\mu}(k)=-ik_{\mu}{\phi}(k),
\end{equation}
which, when substituted into $S_m^{\prime}$, yields the classical equivalence between the two actions, $S_m$ and
$S_m^{\prime}$. Furthermore, variation of $S_m^{\prime}$ with respect to
$F_{\mu}(k)$ leads to the expression of $G^{\mu}(k)$ in terms of $F^{\mu}(k)$,
\begin{equation}
G^{\mu}(k)=F^{\mu}(k)-\frac{1}{2\pi}\int d^2k^{\prime}(g^{{\mu}{\nu}}
+{\epsilon}^{{\mu}{\nu}})
{\lambda}_{{\nu}{\sigma}}(k-k^{\prime})\left[F^{\sigma}(k^{\prime})-
{\epsilon}
^{{\sigma}{\rho}}F_{\rho}(k^{\prime})\right].
\end{equation}
Note that eq.(7) is an integral equation other than an algebraic one that
happens in configuration space, which is
induced by the non-locality of the corresponding Lagrangian
 of eq.(4). At first sight it seems difficult to solve $F^{\mu}(k)$ from
eq.(7). In fact, one can deal with this problem in terms of an algebraic
method shown in the following. In order to avoid solving this integral
equation, one
defines two self-dual tensors as
\begin{eqnarray}
{\cal F}^{\mu}(k) \equiv F^{\mu}(k)-{\epsilon}^{{\mu}{\nu}}F_{\nu}(k),\nonumber
\\
{\cal G}^{\mu}(k) \equiv G^{\mu}(k)-{\epsilon}^{{\mu}{\nu}}G_{\nu}(k),
\end{eqnarray}
and establishes their relationship by using eq.(7),
\begin{equation}
{\cal F}^{\mu}(k)={\cal G}^{\mu}(k).
\end{equation}
We should emphasize that a similar relationship exists in configuration
space for various chiral {\em p}-form actions as pointed out in Ref.[14]. With
the aid of eq.(9), one can easily obtain algebraically from eq.(7)
$F^{\mu}(k)$ expressed in
terms of
$G^{\mu}(k)$,
\begin{equation}
F^{\mu}(k)=G^{\mu}(k)+\frac{1}{2\pi}\int d^2k^{\prime}(g^{{\mu}{\nu}}+{\epsilon}^{{\mu}{\nu}})
{\lambda}_{{\nu}{\sigma}}(k-k^{\prime})\left[G^{\sigma}(k^{\prime})-
{\epsilon}
^{{\sigma}{\rho}}G_{\rho}(k^{\prime})\right].
\end{equation}
We can check from eq.(7) that when the self-duality condition is satisfied,
i.e., ${\cal F}^{\mu}(k)=0$, which is also called an ``on mass shell''
condition,
$F^{\mu}(k)$ and $G^{\mu}(k)$ relate with a duality, $G^{\mu}(k)={\epsilon}
^{{\mu}{\nu}}F_{\nu}(k)$.
Substituting eq.(10) into eq.(5), we obtain the dual action of $S_m$,
\begin{eqnarray}
S_m^{dual}&=& \frac{1}{2}\int d^2k G_{\mu}(k)G^{\mu}(-k)\nonumber\\
& & +\frac{1}{4\pi}\int d^2k d^2k^{\prime}
{\lambda}_{{\mu}{\nu}}(-k-k^{\prime})
\left[G^{\mu}(k)-{\epsilon}
^{{\mu}{\sigma}}G_{\sigma}(k)\right]
\left[G^{\nu}(k^{\prime})-{\epsilon}
^{{\nu}{\rho}}G_{\rho}(k^{\prime})\right]\nonumber\\
& & +\int d^2k {\phi}(k)\left[ik_{\mu}G^{\mu}(-k)\right].
\end{eqnarray}
Variation of eq.(11) with respect to ${\phi}(k)$ gives $ik_{\mu}G^{\mu}(-k)=0$
or $-ik_{\mu}G^{\mu}(k)=0$, whose solution has to be
\begin{equation}
G^{\mu}(k)={\epsilon}^{{\mu}{\nu}}(-ik_{\nu}){\psi}(k) \equiv
{\epsilon}^{{\mu}{\nu}}F_{\nu}[{\psi}(k)],
\end{equation}
where ${\psi}(k)$ is an arbitrary scalar field in momentum space.
Substituting eq.(12) into eq.(11), one finally obtains the dual action in terms
of ${\psi}(k)$,
\begin{eqnarray}
S_m^{dual}=&-&\frac{1}{2}\int d^2k (-ik_{\mu})(ik^{\mu}){\psi}(k){\psi}(-k)
\nonumber\\
&+&\frac{1}{4\pi}\int d^2k d^2k^{\prime}
{\lambda}_{{\mu}{\nu}}(-k-k^{\prime})
(ik^{\mu}-{\epsilon}
^{{\mu}{\sigma}}ik_{\sigma})
({ik^{\prime}}^{\nu}-{\epsilon}
^{{\nu}{\rho}}ik^{\prime}_{\rho}){\psi}(k){\psi}(k^{\prime}).
\end{eqnarray}
This action has the same form as the original one, eq.(4), only with the
replacement of ${\phi}(k)$ by ${\psi}(k)$. As analysed above, ${\phi}(k)$ and
${\psi}(k)$ coincide with each other up to a constant when the self-duality
condition is imposed. Therefore, the {\em k}-space formulation of Siegel's
action is self-dual with respect to ${\phi}(k)$ - ${\psi}(k)$ dualization
expressed by eq.(6) and eq.(12).
\section{Self-duality of the chiral p-form action in D=2(p+1) momentum frame}
By introducing a real {\em p}-form field,
$A_{{\mu}_{1}{\cdots}{\mu}_{p}}(x)$,
we generalize the Siegel action to the $D=2(p+1)$
dimensional space-time,
\begin{eqnarray}
S_c&=&\int d^{D}x \left\{-\frac{1}{2(p+1)!}
{\partial}_{[{\mu}_{1}}A_{{\mu}_{2}
{\cdots}{\mu}_{p+1}]}(x)
{\partial}^{[{\mu}_{1}}A^{{\mu}_{2}
{\cdots}{\mu}_{p+1}]}(x)\right.\nonumber\\
 & & \hspace{13mm}+\frac{1}{2}{{\lambda}^{\mu}}_{\nu}(x)
\left[
{\partial}_{[{\mu}}A_{{\mu}_{1}
{\cdots}{\mu}_{p}]}(x)
-\frac{1}{(p+1)!}{\epsilon}_{{\mu}{\mu}_{1}{\cdots}{\mu}_{p}{\nu}_{1}
{\cdots}
{\nu}
_{p+1}}
{\partial}^{[{\nu}_{1}}A^{{\nu}_{2}
{\cdots}{\nu}_{p+1}]}(x)\right]\nonumber\\
& & \hspace{13mm}\left.{\times} \left[
{\partial}^{[{\nu}}A^{{\mu}_{1}
{\cdots}{\mu}_{p}]}(x)
-\frac{1}{(p+1)!}{\epsilon}^{{\nu}{\mu}_{1}{\cdots}{\mu}_{p}{\sigma}_
{1}
{\cdots}{\sigma}
_{p+1}}
{\partial}_{[{\sigma}_{1}}A_{{\sigma}_{2}
{\cdots}{\sigma}_{p+1}]}(x)\right]\right\}.
\end{eqnarray}
It can be verified that
$A_{{\mu}_{1}{\cdots}{\mu}_{p}}(x)$ indeed describes a chiral {\em p}-form
by following Ref.[8] in which only the $D=2$ and $D=6$ cases are dealt with.
Moreover, the quantization of the theory in higher dimensions
($D=10,14,\cdots$) can be discussed by following Siegel's approach to the
case $D=6$. As pointed out [8] that the generalization is straightforward, a
kind of light-cone gauge should be imposed and then all degrees of freedom
associated with the auxiliary tensor field are eliminated for the cases of
dimensions higher than $D=6$.

As done in the above section, substituting the Fourier
transformations,
\begin{eqnarray}
A_{{\mu}_{1}{\cdots}{\mu}_{p}}(x)&=&\frac{1}{(2\pi)^{D/2}}\int d^{D}k
{e}^{-ik\cdot x}A_{{\mu}_{1}{\cdots}{\mu}_{p}}(k),\nonumber\\
{\lambda}_{{\mu}{\nu}}(x)&=&
\frac{1}{(2\pi)^{D/2}}\int d^{D}k {e}^{-ik\cdot x}{\lambda}_{{\mu}{\nu}}(k),
\end{eqnarray}
into eq.(14), one arrives at Siegel's action in the $D=2(p+1)$ momentum space
spanned by $k_{\mu}$,
\begin{eqnarray}
S_m=&-&\frac{1}{2(p+1)!}\int d^{D}k \left[
{-ik}_{[{\mu}_{1}}A_{{\mu}_{2}
{\cdots}{\mu}_{p+1}]}(k)\right]\left[
{ik}^{[{\mu}_{1}}A^{{\mu}_{2}
{\cdots}{\mu}_{p+1}]}(-k)\right]\nonumber\\
&+&\frac{1}{2(2\pi)^{D/2}}\int d^{D}k d^{D}k^{\prime}
{{\lambda}^{\mu}}_{\nu}(-k-k^{\prime})\nonumber\\
&{\times}& \left[
{ik}_{[{\mu}}A_{{\mu}_{1}
{\cdots}{\mu}_{p}]}(k)
-\frac{1}{(p+1)!}{\epsilon}_{{\mu}{\mu}_{1}{\cdots}{\mu}_{p}{\nu}_{1}
{\cdots}
{\nu}_{p+1}}{ik}^{[{\nu}_{1}}A^{{\nu}_{2}
{\cdots}{\nu}_{p+1}]}(k)\right]\nonumber\\
&{\times}& \left[
{ik^{\prime}}^{[{\nu}}A^{{\mu}_{1}
{\cdots}{\mu}_{p}]}(k^{\prime})
-\frac{1}{(p+1)!}{\epsilon}^{{\nu}{\mu}_{1}{\cdots}{\mu}_{p}{\sigma}_
{1}{\cdots}{\sigma}_{p+1}}
{ik^{\prime}}_{[{\sigma}_{1}}A_{{\sigma}_{2}
{\cdots}{\sigma}_{p+1}]}(k^{\prime})\right].
\end{eqnarray}
Here we note that eq.(16) includes $2D$ momentum integrals. This shows the
non-locality of the corresponding Lagrangian as was seen in the $D=2$ case.

In order to discuss the duality of $S_m$, we introduce two ({\em p}+1)-form
fields in the momentum space,
$F_{{\mu}_{1}{\cdots}{\mu}_{p+1}}(k)$ and
$G_{{\mu}_{1}{\cdots}{\mu}_{p+1}}(k)$, and replace eq.(16) by the following
action,
\begin{eqnarray}
S_m^{\prime}=&-&\frac{1}{2(p+1)!}\int d^{D}kF_{{\mu}_{1}
{\cdots}{\mu}_{p+1}}(k)F^{{\mu}_{1}
{\cdots}{\mu}_{p+1}}(-k)\nonumber\\
&+&\frac{1}{2(2\pi)^{D/2}}\int d^{D}k d^{D}k^{\prime}
{{\lambda}^{\mu}}_{\nu}(-k-k^{\prime})\nonumber\\
&{\times}& \left[F_{{\mu}{\mu}_{1}{\cdots}{\mu}_{p}}(k)
-\frac{1}{(p+1)!}{\epsilon}_{{\mu}{\mu}_{1}{\cdots}{\mu}_{p}{\nu}_{1}
{\cdots}{\nu}_{p+1}}F^{{\nu}_{1}
{\cdots}{\nu}_{p+1}}(k)\right]\nonumber\\
&{\times}& \left[F^{{\nu}{\mu}_{1}{\cdots}{\mu}_{p}}(k^{\prime})
-\frac{1}{(p+1)!}{\epsilon}^{{\nu}{\mu}_{1}{\cdots}{\mu}_{p}{\sigma}_
{1}{\cdots}{\sigma}_{p+1}}F_{{\sigma}_{1}
{\cdots}{\sigma}_{p+1}}(k^{\prime})\right]\nonumber\\
&+&\frac{1}{(p+1)!}\int d^{D}kG^{{\mu}_{1}
{\cdots}{\mu}_{p+1}}(-k)\left[F_{{\mu}_{1}{\cdots}{\mu}_{p+1}}(k)+
ik_{[{\mu}_{1}}A_{{\mu}_{2}{\cdots}{\mu}_{p+1}]}(k)\right],
\end{eqnarray}
where
$F_{{\mu}_{1}{\cdots}{\mu}_{p+1}}(k)$ and
$G_{{\mu}_{1}{\cdots}{\mu}_{p+1}}(k)$ act,
at present, as independent auxiliary fields. Similar to the $D=2$ case, the
third term can be obtained by substituting the Fourier transformations of the
fields, {\em i.e.}, eq.(15), into the following term,
$$
\frac{1}{(p+1)!}\int d^{D}xG^{{\mu}_{1}{\cdots}{\mu}_{p+1}}(x)\left[
F_{{\mu}_{1}{\cdots}{\mu}_{p+1}}(x)-{\partial}_{[{\mu}_{1}}
A_{{\mu}_{2}{\cdots}{\mu}_{p+1}]}(x)\right].
$$
Variation of
eq.(17) with respect to $G^{{\mu}_{1}{\cdots}{\mu}_{p+1}}(-k)$ gives
\begin{equation}
F_{{\mu}_{1}{\cdots}{\mu}_{p+1}}(k)=
-ik_{[{\mu}_{1}}A_{{\mu}_{2}{\cdots}{\mu}_{p+1}]}(k),
\end{equation}
which yields the equivalence between the actions, eq.(16) and eq.(17). On the
other hand, variation of eq.(17) with respect to
$F_{{\mu}_{1}{\cdots}{\mu}_{p+1}}(k)$ leads to the expression of
$G^{{\mu}_{1}{\cdots}{\mu}_{p+1}}(k)$ in terms of
$F^{{\mu}_{1}{\cdots}{\mu}_{p+1}}(k)$,
\begin{eqnarray}
G^{{\mu}_{1}{\cdots}{\mu}_{p+1}}(k)&=&
F^{{\mu}_{1}{\cdots}{\mu}_{p+1}}(k)-\frac{1}{(2\pi)^{D/2}}\int d^{D}k^{\prime}{\lambda}_{{\mu}{\nu}}(k-k^{\prime})\nonumber\\
 & &{\times}\left[g^{{\mu}[{\mu}_{1}}{\cal F}^{{\mu}_{2}{\cdots}{\mu}_{p+1}]{\nu}}(k^{\prime})
+{\epsilon}^{{\mu}_{1}{\cdots}{\mu}_{p+1}{\mu}{\nu}_{1}{\cdots}
{\nu}_{p}}{{\cal F}^{\nu}}_{{\nu}_{1}
{\cdots}{\nu}_{p}}(k^{\prime})\right],
\end{eqnarray}
where ${\cal F}^{{\mu}_{1}{\cdots}{\mu}_{p+1}}(k)$
is defined as the difference of the field strength $F^{{\mu}_{1}{\cdots}{\mu}_{p+1}}(k)$ and its Hodge dual,
\begin{equation}
{\cal F}^{{\mu}_{1}{\cdots}{\mu}_{p+1}}(k) \equiv
F^{{\mu}_{1}{\cdots}{\mu}_{p+1}}(k)
-\frac{1}{(p+1)!}{\epsilon}^{{\mu}_{1}{\cdots}{\mu}_{p+1}{\nu}_{1}
{\cdots}{\nu}_{p+1}}F_{{\nu}_{1}{\cdots}{\nu}_{p+1}}(k),
\end{equation}
which is also called the self-dual tensor.
Note that eq.(19) is an integral equation. In order to obtain algebraically $F^{{\mu}_{1}{\cdots}{\mu}_{p+1}}(k)$ in terms of $G^{{\mu}_{1}{\cdots}{\mu}_{p+1}}(k)$ without
solving the integral equation, as was done in the last section, one defines another field strength difference/self-dual tensor which is relevant to $G^{{\mu}_{1}{\cdots}{\mu}_{p+1}}(k)$,
\begin{equation}
{\cal G}^{{\mu}_{1}{\cdots}{\mu}_{p+1}}(k) \equiv
G^{{\mu}_{1}{\cdots}{\mu}_{p+1}}(k)
-\frac{1}{(p+1)!}{\epsilon}^{{\mu}_{1}{\cdots}{\mu}_{p+1}{\nu}_{1}
{\cdots}{\nu}_{p+1}}G_{{\nu}_{1}{\cdots}{\nu}_{p+1}}(k),
\end{equation}
and then establishes the relationship between these two self-dual tensors by using eq.(19),
\begin{equation}
{\cal F}^{{\mu}_{1}{\cdots}{\mu}_{p+1}}(k)={\cal G}^{{\mu}_{1}{\cdots}{\mu}_{p+1}}(k),
\end{equation}
whose similar form,
as mentioned above, exists in various chiral {\em p}-forms in
configuration space [14].  With eq.(22), one can algebraically invert eq.(19) and obtain $F^{{\mu}_{1}{\cdots}{\mu}_{p+1}}(k)$ expressed in terms of $G^{{\mu}_{1}{\cdots}{\mu}_{p+1}}(k)$,
\begin{eqnarray}
F^{{\mu}_{1}{\cdots}{\mu}_{p+1}}(k)&=&
G^{{\mu}_{1}{\cdots}{\mu}_{p+1}}(k)+\frac{1}{(2\pi)^{D/2}}\int d^{D}k^{\prime}{\lambda}_{{\mu}{\nu}}(k-k^{\prime})\nonumber\\
 & &{\times}\left[g^{{\mu}[{\mu}_{1}}{\cal G}^{{\mu}_{2}{\cdots}{\mu}_{p+1}]{\nu}}(k^{\prime})
+{\epsilon}^{{\mu}_{1}{\cdots}{\mu}_{p+1}{\mu}{\nu}_{1}{\cdots}
{\nu}_{p}}{{\cal G}^{\nu}}_{{\nu}_{1}
{\cdots}{\nu}_{p}}(k^{\prime})\right].
\end{eqnarray}
We can verify from eq.(19) that when the self-duality condition is satisfied,
i.e.,
$${\cal F}^{{\mu}_{1}{\cdots}{\mu}_{p+1}}(k)=0,$$
$F^{{\mu}_{1}{\cdots}{\mu}_{p+1}}(k)$ and $G^{{\mu}_{1}{\cdots}{\mu}_{p+1}}(k)$ relate with a duality,
$$G^{{\mu}_{1}{\cdots}{\mu}_{p+1}}(k)=
\frac{1}{(p+1)!}{\epsilon}^{{\mu}_{1}{\cdots}{\mu}_{p+1}{\nu}_{1}
{\cdots}{\nu}_{p+1}}F_{{\nu}_{1}{\cdots}{\nu}_{p+1}}(k).
$$
Now substituting eq.(23) into the action, eq.(17), and making tedious
calculations, we obtain the dual Siegel action in the $D=2(p+1)$ momentum
space,
\begin{eqnarray}
S_m^{dual}&=&\frac{1}{2(p+1)!}\int d^{D}kG_{{\mu}_{1}
{\cdots}{\mu}_{p+1}}(k)G^{{\mu}_{1}
{\cdots}{\mu}_{p+1}}(-k)\nonumber\\
& &+\frac{1}{2(2\pi)^{D/2}}\int d^{D}k d^{D}k^{\prime}
{\lambda}_{{\mu}{\nu}}(-k-k^{\prime})
{\cal G}^{{\mu}{\mu}_{1}{\cdots}{\mu}_{p}}(k){{\cal G}^{\nu}}_{{\mu}_{1}{\cdots}{\mu}_{p}}(k^{\prime})\nonumber\\
& &+\int d^{D}kA_{{\mu}_{1}{\cdots}{\mu}_{p}}(k)\left[
ik_{\mu}G^{{\mu}{\mu}_{1}{\cdots}{\mu}_{p}}(-k)\right].
\end{eqnarray}
Variation of eq.(24) with respect to $A_{{\mu}_{1}
{\cdots}{\mu}_{p}}(k)$ gives $ik_{\mu}G^{{\mu}{\mu}_{1}{\cdots}{\mu}_{p}}(-k)=0$,
 whose solution has to be
\begin{eqnarray}
G^{{\mu}_{1}{\cdots}{\mu}_{p+1}}(k)&=&
\frac{1}{(p+1)!}{\epsilon}^{{\mu}_{1}{\cdots}{\mu}_{p+1}{\nu}_{1}
{\cdots}{\nu}_{p+1}}\left[-ik_{[{\nu}_{1}}B_{{\nu}_{2}
{\cdots}{\nu}_{p+1}]}(k)\right]\nonumber\\
&{\equiv}&
\frac{1}{(p+1)!}{\epsilon}^{{\mu}_{1}{\cdots}{\mu}_{p+1}{\nu}_{1}
{\cdots}{\nu}_{p+1}}F_{{\nu}_{1}{\cdots}{\nu}_{p+1}}[B(k)],
\end{eqnarray}
where $B_{{\nu}_{1}{\cdots}{\nu}_{p}}(k)$
is an arbitrary {\em p}-form field. When eq.(25) is
substituted into the dual action, eq.(24), one finally obtains the result that the dual action is the same as the original action, eq.(16), only with the replacement of $A_{{\nu}_{1}
{\cdots}{\nu}_{p}}(k)$ by $B_{{\nu}_{1}{\cdots}{\nu}_{p}}(k)$. Consequently, the {\em k}-space formulation of
Siegel's action in $D=2(p+1)$ dimensions is self-dual with respect to
$A_{{\nu}_{1}{\cdots}{\nu}_{p}}(k)-B_{{\nu}_{1}{\cdots}{\nu}_{p}}(k)$ dualization given by eq.(18) and eq.(25).
\section{Conclusion}
In this note we have generalized Siegel's model to the $D=2(p+1)$ dimensional
space-time, and have extended for this model duality investigations from the
configuration frame to the momentum frame and hence uncovered its self-duality
with respect to dualization of chiral fields in the
momentum space. The characteristic that does not exist in configuration space
is duality investigation of non-local Lagrangians and algebraic solution of
integral equations. Here we emphasize that the introduction of two self-dual
tensors and the establishment of their relationship are crucial
in realizing the whole procedure of investigation.

The PST action [11,13] is not polynomial, however, the
non-polynomiality can be formally eliminated in terms of a re-definition of
auxiliary fields. Here we take
the two dimensional case as our example. (Generalization to higher
dimensions is straightforward. See Ref.[12].) The PST action reads
\begin{eqnarray*}
S&=& \int d^{2}x\bigg\{-\frac{1}{2}{\partial}_{\mu}{\phi}(x){\partial}^{\mu}
{\phi}(x)\\
& & +\frac{1}
{2{\partial}_{\eta}a(x){\partial}^{\eta}a(x)}
{\partial}_{\mu}a(x)[{\partial}^{\mu}{\phi}(x)-{\epsilon}^{{\mu}{\sigma}}
{\partial}_{\sigma}{\phi}(x)]
{\partial}_{\nu}a(x)[{\partial}^{\nu}{\phi}(x)-{\epsilon}^{{\nu}{\rho}}
{\partial}_{\rho}{\phi}(x)]\bigg\},
\end{eqnarray*}
where $a(x)$ is an auxiliary  scalar field.
If we introduce an auxiliary tensor field ${\lambda}_{{\mu}{\nu}}(x)$
which is defined
by

\[
{\lambda}_{{\mu}{\nu}}(x)\equiv \frac{{\partial}_{\mu}a(x){\partial}_{\nu}a(x)}
{{\partial}_{\eta}a(x)
{\partial}^{\eta}a(x)},
\]
\noindent
the PST action appears then in the form of the Siegel action, eq.(2).
In this sense, we may say that the duality investigation of the PST action in
the momentum space is the same as that of Siegel's action formally. We would
like to thank the referee for pointing out this  {\em
solely  formal equivalence}.
However, our treatment is valid for any chiral $p$-form action  in which
polynomial or specifically Lagrange-multiplier auxiliary fields are
introduced. This validity has nothing to do with gauge symmetries of these
actions, that is, our treatment works whether the actions contain first-class
or second-class constraints. The property appears quite naturally when one
applies the whole procedure of our duality investigation to a gauge invariant
(first-class constraint) model, for instance, to the McClain-Wu-Yu action
[10]. As to the {\em direct} duality investigation of the PST action in the
momentum space, difficulties exist {\em not due to} the first-class
constraints, {\em but} the non-polynomiality of the auxiliary fields. To deal
with the problem requires us to develop an appropriate technique of the
Fourier transformation. This topic is beyond the context of the present note,
and we shall consider it later in a separate work.

Finally, we make a few remarks upon the non-locality of Siegel's Lagrangian in
the momentum frame.

(i) As we know, the non-locality originates from the non-linearity which
is imposed by the manifest Lorentz invariance of Siegel's action.
Therefore,
we conclude that the manifest Lorentz invariance of actions in configuration
space, which is a quite commonly required property of actions, sometimes
produces the non-locality of corresponding Lagrangians in momentum space.

(ii)
In Ref.[4] a non-local Lagrangian of a chiral 0-form was proposed and then a
local fermionic Lagrangian was proved to be equivalent to the non-local
bosonic one in terms of chiral bosonization in configuration space. In our
case we have obtained the non-local Lagrangian eq.(4) in momentum space. We
may ask whether or not its local and/or equivalent fermionic formulation
exists. In addition, we may need to develop a technique of chiral bosonization
in momentum space.

(iii)
A problem which is closely related to (ii) is the quantization of non-local
Lagrangians in momentum space. For instance, we may generalize Dirac's method
[17] which has been treated as a standard approach of quantization. To this
end, we have to deal with several basic concepts, such as the definition of
canonical momenta, the classification of first and second class constraints,
and the construction of Dirac brackets, and so on.
This is now under consideration.
\vskip 38pt
\noindent
{\bf Acknowledgments}
\par
Y.-G. Miao acknowledges supports by an Alexander von Humboldt fellowship, by
the National Natural Science Foundation
of China under grant No.19705007, and by the Ministry of Education of China
under the special project for scholars returned from abroad. D.K. Park
acknowledges support from the Basic Research Program of the Korea Science
and Engineering Foundation (Grant No. 2001-1-11200-001-2).

\end{document}